\documentclass[prl,aps,twocolumn,floatfix,longbibliography,superscriptaddress]{revtex4-2}
\usepackage{hyperref}
\usepackage[utf8]{inputenc}
\usepackage{natbib}
\usepackage{graphicx}
\usepackage{color}
\usepackage{amsmath}
\usepackage{amssymb}
\usepackage{gensymb}
\usepackage{float}

\begin{document}

\title{Pair density modulation from glide symmetry breaking and nematic superconductivity}

\author{Micha{\l} Papaj}
\affiliation{Department of Physics and Texas Center for Superconductivity at the University of Houston (TcSUH), Houston, TX 77204, USA}

\author{Lingyuan Kong}
\affiliation{T. J. Watson Laboratory of Applied Physics, California Institute of Technology, 1200 East California Boulevard, Pasadena, California 91125, USA}
\affiliation{Institute for Quantum Information and Matter, California Institute of Technology, Pasadena, California 91125, USA}

\author{Stevan Nadj-Perge}
\affiliation{T. J. Watson Laboratory of Applied Physics, California Institute of Technology, 1200 East California Boulevard, Pasadena, California 91125, USA}
\affiliation{Institute for Quantum Information and Matter, California Institute of Technology, Pasadena, California 91125, USA}

\author{Patrick A. Lee}
\affiliation{Department of Physics, Massachusetts Institute of Technology, Cambridge, MA 02139, USA}

\begin{abstract}
Pair density modulation is a superconducting state, recently observed in exfoliated iron-based superconductor flakes, in which the superconducting gap oscillates strongly with the same periodicity as the underlying crystalline lattice. We propose a microscopic model that explains this modulation through a combination of glide-mirror symmetry breaking and the emergence of nematic superconductivity. The first ingredient results in a sublattice texture on the Fermi surface, which is aligned with the anisotropic superconducting gap of the nematic $s_\pm+d$ state. This gives rise to distinctive gap maxima and minima located on the two inequivalent iron sublattices while still being a zero-momentum pairing state. We discuss how further investigation of such modulations can give insight into the nature of the superconducting pairing, such as the signs of the order parameters and visualization of a phase transition to a mixed two-component state using local probes.
\end{abstract}

\maketitle

\textit{Introduction}.---
Modulated superconducting orders quickly attracted significant attention after the first proposals by Fulde, Ferrell, Larkin, and Ovchinnikov (FFLO) \cite{larkinNonuniformStateSuperconductors1964a, fuldeSuperconductivityStrongSpinExchange1964a}. Although the FFLO states required explicit time-reversal symmetry breaking, further research revealed that the superconducting order parameters that vary periodically in space can also arise as a result of strong correlations. Such states are called pair density waves (PDW) \cite{agterbergPhysicsPairDensityWaves2020} and have been proposed to exist in many classes of materials, such as cuprates \cite{hamidianDetectionCooperpairDensity2016, ruanVisualizationPeriodicModulation2018, duImagingEnergyGap2020a, wangScatteringInterferenceSignature2021, chenIdentificationNematicPair2022}, iron-based superconductors \cite{zhaoSmecticPairdensitywaveOrder2023, liuPairDensityWave2023}, kagome materials \cite{chenRotonPairDensity2021, dengChiralKagomeSuperconductivity2024}, and others \cite{liuDiscoveryCooperpairDensity2021, guDetectionPairDensity2023, weiDiscoverySmecticCharge2024}. In most PDW states, the modulation period is not directly related to the crystalline unit cell dimensions, and so the emergence of PDW order results in breaking the translational symmetry of the crystal. The usually observed wave lengths are on the order of several nanometers, spanning multiple unit cells. Moreover, while in many PDW cases the theory predicts gapless excitations due to formation of Fermi arcs at the Fermi energy \cite{baruchSpectralSignaturesModulated2008, bergStripedSuperconductorsHow2009, leeAmpereanPairingPseudogap2014}, in experiments the modulation is usually imposed on top of a much larger overall superconducting gap structure, suggesting coexistence of PDW and conventional zero-momentum pairing. However, this can be difficult to reconcile as the two phases are usually strongly competing and the superconducting susceptibility divergence at $\mathbf{q}=0$ usually dominates over finite-$\mathbf{q}$ states. In many cases, questions also arise whether the PDW is a primary order, or a secondary one coming from an underlying normal state modulation, such as charge density wave states, or even due to pair-breaking scattering that can induce gap oscillations \cite{gaoPairbreakingScatteringInterference2024}.

However, recently a different type of modulated superconducting order was observed in exfoliated flakes of FeTe$_{0.55}$Se$_{0.45}$ \cite{kongCooperpairDensityModulation2025}, which is called pair density modulation (PDM). The PDM experiment exhibited several differences from previous measurements of PDW: modulation preserves lattice translational symmetry, since it is periodic with the two-iron unit cell of the material, the modulation amplitude was much larger than other cases observed so far (reaching 30~\% of the average gap value), and the gap extrema were located on the two iron sublattices, with the role of the inequivalent iron sites exchanged between short-range domains. Nematic distortion, a frequent phenomenon in iron-based compounds \cite{fernandesWhatDrivesNematic2014, bohmerNematicityNematicFluctuations2022a, fernandesIronPnictidesChalcogenides2022, mukasaHighpressurePhaseDiagrams2021}, was observed in the quasiparticle interference spectra, leading to questions about the origin of the modulation and possible information that can be extracted from it, which should be answered by a microscopic theory.

In this work, we present such a theory based on the combination of glide-mirror symmetry breaking and the appearance of mixed $s_\pm+d$ superconducting order previously studied in iron-based superconductors \cite{livanasNematicityMixed2015, chenNematicitySuperconductivityCompetition2020, kangSuperconductivityFeSeRole2018}. Although each of these ingredients on its own will not lead to any spatial modulation, the combination of the Fermi surface sublattice texture (which was also proposed to be important for modulations in kagome superconductors \cite{schwemmerSublatticeModulatedSuperconductivity2024}) and superconducting (SC) order parameter anisotropy in momentum space results in gap oscillations with maxima and minima at the two inequivalent iron sites. This mechanism allows for a large modulation of the gap, is not reliant on any charge density wave in the normal state, as it arises purely through a superconducting transition, and requires only zero-momentum pairing, which should be preferred in a wide range of circumstances. We also predict that generically the superconducting transition temperature and the temperature at which the modulation appears are distinct, enabling visualization of the mixed SC order emergence using local probes. Our proposal can thus serve to gather new insight into the mechanisms behind superconducting pairing in iron-based superconductors and other materials.

\begin{figure}
    \centering
    \includegraphics[width=0.99\columnwidth]{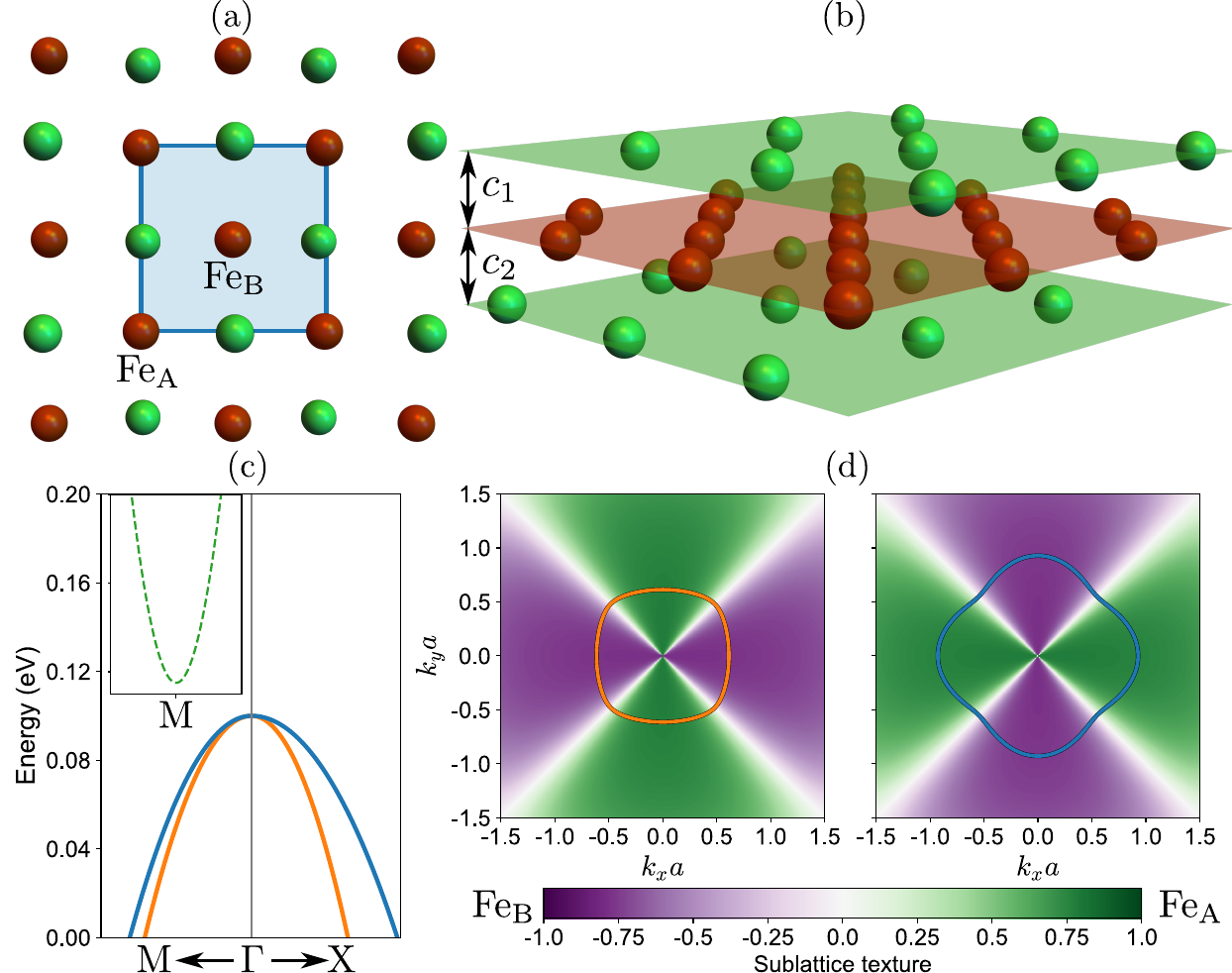}
    \caption{(a) Top view of the two-iron unit cell (blue square) with Fe atoms (brown) and Se/Te atoms (green). The Se site that is nearest neighbor to the Fe$_\text{A}$ site along the $x$ axis is above the Fe plane while that along $y$ is below the plane. (b) Side view of the unit cell, which indicates the distance between the Fe plane and Se planes ($c_1$ for upper and $c_2$ for lower). When $c_1\neq c_2$, the glide symmetry is broken. (c) The dispersion of the two lower energy bands of the continuum model (blue and orange) in the vicinity of $\Gamma$ point in the direction of M and X points. The inset shows the M point band located above the Fermi energy (green dashed line). (d) The sublattice texture of the two $\Gamma$ bands that cross the Fermi energy. States on the Fermi surface in the green region are more localized on Fe$_\text{A}$, while states in purple region are localized on Fe$_\text{B}$.}
    \label{fig:model}
\end{figure}

\textit{Model}.---
We begin by constructing a tight-binding model using the Koster-Slater approach \cite{slaterSimplifiedLCAOMethod1954, takegaharaSlaterKosterTablesElectrons1980, moreoPropertiesTwoorbitalModel2009, daghoferThreeOrbitalModel2010} that describes the full unit cell of Fe(Se, Te) with two Fe and two Se/Te atoms as shown in Fig.~\ref{fig:model}(a). As we are interested in the low-energy properties of the material close to the Fermi surface, we choose to only consider $d_{xz}$, $d_{yz}$, and $d_{xy}$ orbitals on each of the iron atoms, since these are the most relevant to the pockets around the $\Gamma$ and M points. Each of the Se/Te atoms contributes $p_x$, $p_y$, and $p_z$ orbitals. Since the observed PDM reveals a stark difference between the density of states in the superconducting state at the Fe$_\text{A}$ and Fe$_\text{B}$ sublattices while maintaining the translational symmetry of the two-iron unit cell, this means that the glide symmetry of the original crystal structure must be broken. We take this into account by assuming that the upper and lower Se/Te planes are no longer at the same distance from the Fe plane. This is captured by $c_1 \neq c_2$ with the distances $c_1, c_2$ as defined in Fig.~\ref{fig:model}(b). This choice of symmetry breaking is motivated by the radical change in the $c$-axis lattice constant of the exfoliated flakes \cite{kongCooperpairDensityModulation2025}, together with the proximity of the flake surface inherent to the STM measurements and the random distribution of Se and Te in the alloy that affects their vertical position.

Therefore, in our model, the atoms in the unit cell are located at: Fe$_\text{A}$ - (0, 0, 0), Fe$_\text{B}$ - ($\frac{a}{2}, \frac{a}{2}$, 0), Se$_\text{A}$ - ($\frac{a}{2}$, 0, $c_1$), Se$_\text{B}$ - (0, $\frac{a}{2}$, $-c_2$), where the lattice vectors are ($a$, 0, 0) and (0, $a$, 0) with $a$ being the distance between iron atoms that belong to the same sublattice (the next-nearest neighbors). With such a setup, we find all the possible hoppings between Fe and Se atoms in terms of the Koster-Slater integrals. We then perform a Schrieffer-Wolff transformation \cite{winklerSpinOrbitCoupling2003, bravyiSchriefferWolffTransformation2011} to treat the Se $p$ orbitals in the second order of perturbation theory, arriving at a tight-binding Hamiltonian in the iron sublattice and orbital space (see the Supplemental Material for the full form \cite{SM}). As the experiment indicates that the exfoliated flakes contain only hole-like Fermi surfaces, based on the fact that in the majority of the iron-based superconductors the hole pockets are localized close to the $\Gamma$ point, we further simplify the model by performing a Taylor expansion of the tight-binding model in the vicinity of $k=0$. This is similar in spirit to previously derived continuum models that explicitly keep the orbital structure of the electronic states \cite{cvetkovicSpaceGroupSymmetry2013, christensenSpinReorientationDriven2015, fernandesLowenergyMicroscopicModels2016}. Here, however, due to the glide symmetry breaking we also keep the sublattice degree of freedom in the model. Keeping terms up to second order in $k$ and only $d_{xz}, d_{yz}$ orbitals, we arrive at the following continuum model:

\begin{equation}
    \label{eq:continuum_Hamiltonian}
    H_0(\mathbf{k}) = 
    \begin{pmatrix}
        \epsilon_{\text{A}} & \epsilon_{\text{A},12} & \epsilon_{\text{AB},1} & \epsilon_{\text{AB},12} \\
        \epsilon_{\text{A},12} & \epsilon_{\text{A}} & -\epsilon_{\text{AB},12} & \epsilon_{\text{AB},2} \\
        \epsilon_{\text{AB},1} & -\epsilon_{\text{AB},12} & \epsilon_{\text{B}} & \epsilon_{\text{B},12} \\
        \epsilon_{\text{AB},12} & \epsilon_{\text{AB},2} & \epsilon_{\text{B},12} & \epsilon_{\text{B}}
    \end{pmatrix}
\end{equation}
with $\epsilon_{\text{A/B}} = -\alpha k^2 \mp \beta (k_x^2-k_y^2) - \mu$, $\epsilon_{\text{AB},1/2} = \xi (k^2 - 8) \pm \rho k_x k_y$, $\epsilon_{\text{A/B},12}= \pm \delta + \gamma (k_x^2-k_y^2) \pm \kappa (k^2 - 4)$, $\epsilon_{\text{AB},12}=\chi k_x k_y$, where $\alpha, \beta, \delta, \gamma, \kappa, \xi, \rho, \chi$ are the parameters of the model, 1 and 2 indicate $d_{xz}$ and $d_{yz}$ orbitals, respectively, and $\mu$ is the chemical potential. The quasiparticle interference measurements also indicate the presence of electron-like band close to the Fermi energy. As electron pockets most often occur close to the M points of the two-iron Brillouin zone, we include in our model an additional parabolic band, which dispersion for small momenta close to $(\pi, \pi)$ is given by $\epsilon_M = \epsilon_0 + \nu k^2 - \mu$. To simplify the model, we ignore this band's sublattice and orbital structure, which would normally contain a strong admixture of $d_{xy}$ orbitals.

The spectrum of this model consists of four bands which are split into two doublets at the $\Gamma$ point. We are interested in the lower pair of bands that is close to the Fermi energy. Fig.~\ref{fig:model}(c) shows an example low-energy dispersion of the model (full spectrum shown in SM), together with the band located around the M point indicated by the dashed line displayed in the inset. The key feature of this model is the appearance of sublattice texture when glide-mirror symmetry is broken. This means that after symmetry breaking, some parts of the Fermi surface contain states which are more strongly localized on one sublattice than on the other. We define the texture operator as $\tau_z$, which is the Pauli matrix for the sublattice degree of freedom. The two hole-like Fermi pockets together with their sublattice textures are shown in Fig.~\ref{fig:model}(d). We can see that close to the $\Gamma$ point, sublattice texture has $d$-like symmetry, that is, it switches sign (and correspondingly, the sublattice) after $\pi/2$ rotation in momentum space.

\begin{figure}
    \centering
    \includegraphics[width=0.999\columnwidth]{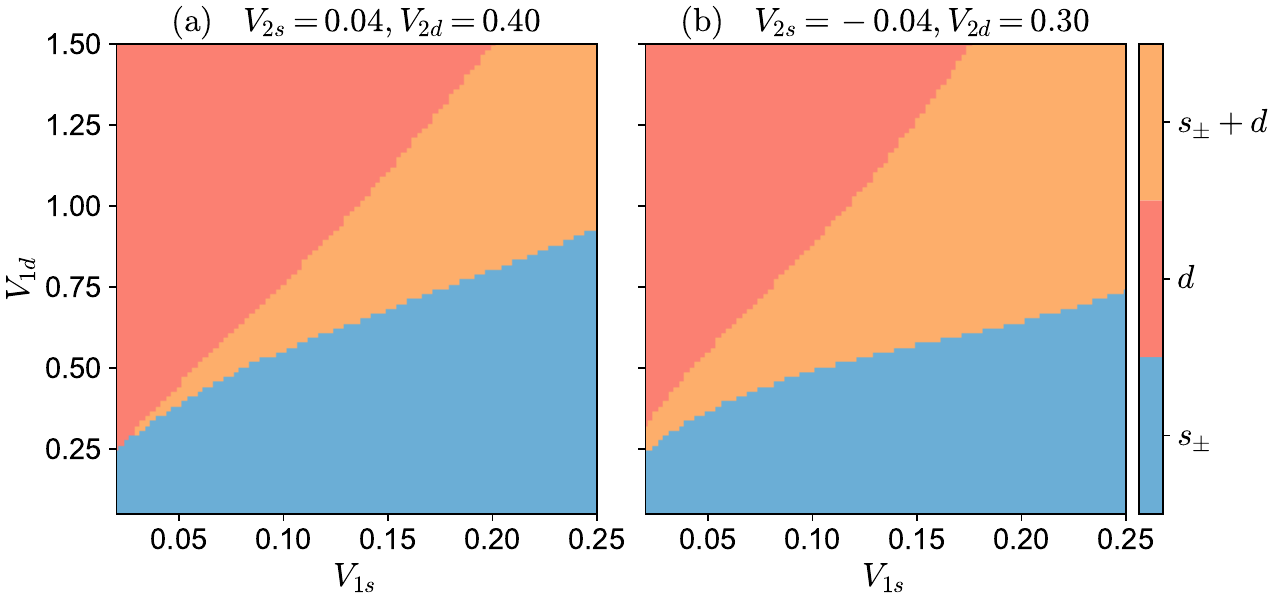}
    \caption{Examples of zero temperature phase diagrams for the order parameters of one of the bands for (a) purely attracitve interaction and (b) for repulsive interaction in $s_\pm$ channel. In both cases, there exists a wide region of interaction strengths for which mixed $s_\pm+d$ order parameter emerges.}
    \label{fig:phases}
\end{figure}

We can now consider the interactions between particles with energies close to the Fermi level. To keep our discussion fairly general, we study intra- and interband interactions of separable form in the $s_\pm$ channel, which is the most commonly suggested pairing for iron-based superconductors \cite{mazinUnconventionalSuperconductivitySign2008, hanaguriUnconventionalSWaveSuperconductivity2010, chenDirectVisualizationSignreversal2019}, and the $d$ channel. The form of interaction itself is BCS-like with pairs of particles of opposite momentum and spin:
\begin{equation}
\begin{split}
    H_\text{int} = -\frac{1}{N} \sum_{\mathbf{k}, \mathbf{k'}} \left[  V_1(\mathbf{k}, \mathbf{k'}) \sum_i c^\dagger_{i\mathbf{k}\uparrow} c^\dagger_{i-\mathbf{k}\downarrow} c_{i-\mathbf{k'}\downarrow} c_{i\mathbf{k'}\uparrow} + \right. \\
    \left.  V_2(\mathbf{k}, \mathbf{k'}) \sum_{i\neq j} c^\dagger_{i\mathbf{k}\uparrow} c^\dagger_{i-\mathbf{k}\downarrow} c_{j-\mathbf{k'}\downarrow} c_{j\mathbf{k'}\uparrow} \right]
\end{split} 
\end{equation}
where $i,j$ indices go over the three bands of the model, $c^\dagger_{i\mathbf{k}\uparrow/\downarrow}$ are the corresponding creation operators for the particles in the band $i$ of spin up/down, $V_n(\mathbf{k}, \mathbf{k'}) = V_{ns}f_s(\mathbf{k})f_s(\mathbf{k'}) + V_{nd}f_d(\mathbf{k})f_d(\mathbf{k'})$ and $f_s(\mathbf{k}) = \cos(k_x) + \cos(k_y)$, $f_d(\mathbf{k}) = \cos(k_x) - \cos(k_y)$ are the $s_\pm$ and $d$ channel interaction form factors, respectively. We therefore characterize the interaction strength by four parameters for intra- and interband $s_\pm$ and $d$ channels. In principle, the interaction strengths will be different for each band (and band pair for interband terms), but we choose to limit our analysis to the four parameters described above for simplicity.

\textit{Mean-field SC order parameters and PDM origin}.---
We then perform mean-field self-consistent calculations to find the superconducting order parameters for all three of the bands at zero temperature. Since we allow for interaction strengths to be different for intra- and interband processes, in general this means that we can have six order parameters, of $s_\pm$ and $d$ nature for each of the three bands:
\begin{equation}
    \Delta^{(i)}(\mathbf{k}) = \Delta_{s}^{(i)} f_s(\mathbf{k}) + \Delta_{d}^{(i)} f_d(\mathbf{k})
\end{equation}
The solutions of the gap equations (more details in the Supplemental Material \cite{SM}) can then be categorized into three classes: pure $s_\pm$-wave, pure $d$-wave, and mixed $s_\pm + d$ phase. The regions of stability of these phases at zero temperature depend on the intra- and interband interaction strengths, as well as the chemical potential. Examples of phase diagrams are presented in Fig.~\ref{fig:phases}. We observe that there are large regions of the parameter space in which the mixed phase exists, which are stable to changes in chemical potential and variations in interaction strengths. Due to the inclusion of interband interactions, while the exact values of the SC order parameter components are different between the bands, the phase boundaries are generally the same for each of the bands \cite{SM}. However, there are some cases for which the order parameters in each of the bands belong to a different phase. Moreover, the interband interaction is necessary for the mixed phase to appear, even if its strength is much weaker than that of the intraband part. Without it, we observe a first-order phase transition between pure $s_\pm$ and $d$ phases as a function of the relative interaction strength in the two channels.

The above conclusions are also independent of whether the interband interaction is attractive [Fig.~\ref{fig:phases}(a)] or repulsive [Fig.~\ref{fig:phases}(b)], provided that that there exists a sufficiently strong attractive interaction in some channel. What changes, however, is the relative sign between the order parameters on the two bands. The order parameters are in phase for attractive intra- and interband interactions, but there exists a region of the phase diagram where the signs can become opposite. An example of such situation is when $V_{1s} > 0$ and $V_{2s} < 0$, which results in $\Delta_{s}^{(1)} < 0$ and $\Delta_{s}^{(2)} > 0$. In this scenario, the signs of the $d$-wave order parameters remain synchronized.

\begin{figure}
    \centering
    \includegraphics[width=0.99\columnwidth]{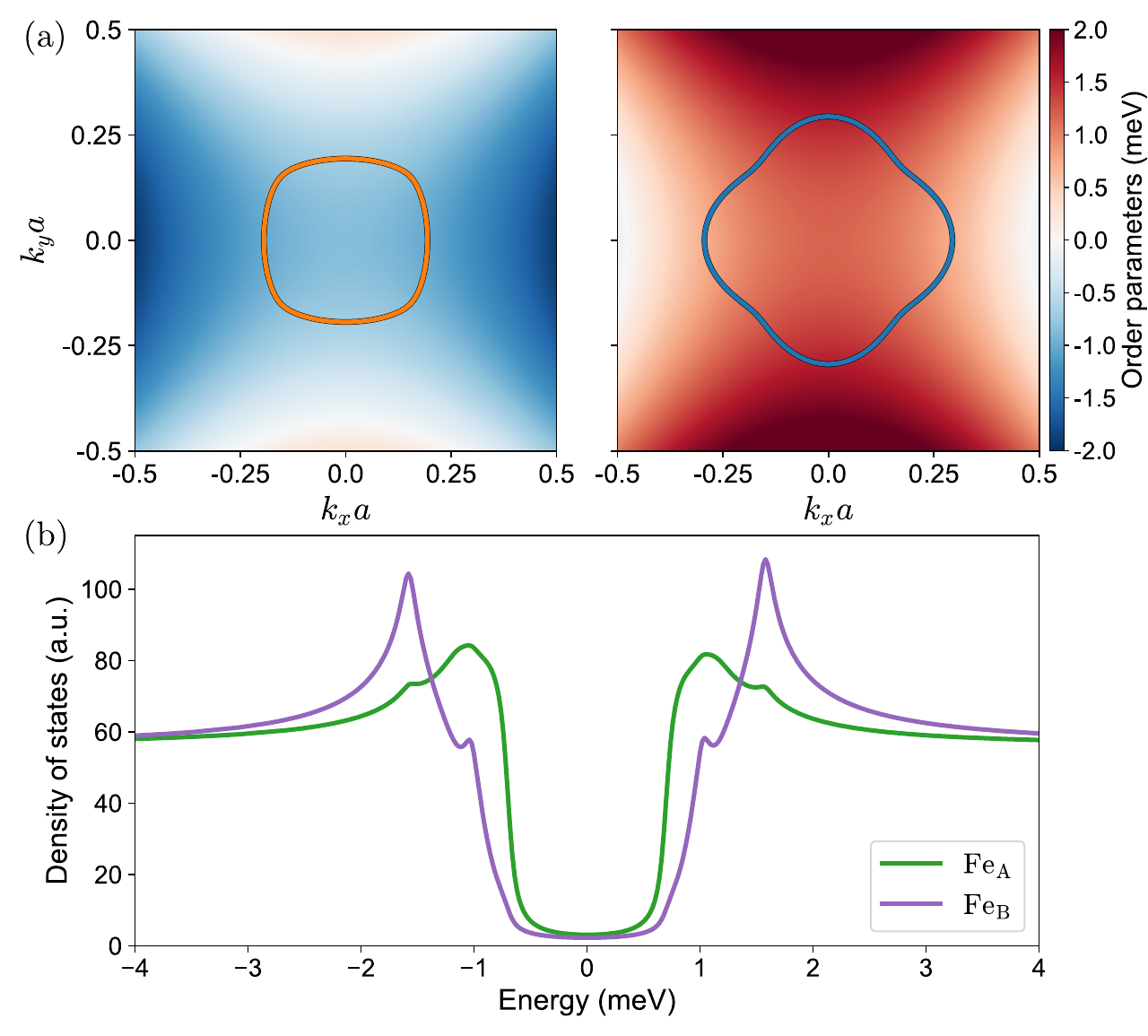}
    \caption{(a) The superconducting order parameters for the two bands obtained for $V_{1s}=0.034$, $V_{1d}=0.64$, $V_{2s}=-0.017$, $V_{2d}=0.32$. The nodal line of the order parameter for either band is not crossing the Fermi surface, so the spectrum will remain fully gapped. b) The density of states at the two iron sublattices, which demonstrates that the position of the coherence peaks shifts when performing measurements at Fe$_\mathrm{A}$ and Fe$_\mathrm{B}$ sites.}
    \label{fig:dos}
\end{figure}

To visualize this, we plot the momentum space distribution of the order parameters of both bands in Fig.~\ref{fig:dos}(a) for a system in the mixed $s_\pm+d$ phase with repulsive interband interactions $V_{2s} < 0$. We see that the band-1 order parameter is negative over the entire Fermi surface, while the band-2 order parameter is positive for the whole contour. At the same time, this means that the entire dispersion will be fully gapped, since the nodal line of both order parameters is far away from where the Fermi surfaces lie. Nevertheless, there is a strong anisotropy of the order parameters when traced around the Fermi pockets. With the particular choice of parameters shown in Fig.~\ref{fig:dos}, this means that the magnitude of the order parameter for band 1 is the largest for parts of the Fermi surface where $k_y \approx 0$ and the smallest where $k_x \approx 0$. The situation is reversed for band 2.

This order parameter anisotropy is the key for the emergence of pair density modulation. To understand this connection, we have to correlate the order parameter map with the sublattice texture of Fig.~\ref{fig:model}(d). In doing so, we notice that for both bands the momentum space region with the largest SC order parameter coincides with the part of the Fermi surface with states largely localized on the Fe$_\mathrm{B}$ sublattice and the smallest order parameter with states on the Fe$_\mathrm{A}$ sublattice. This directly translates to a larger superconducting gap visible on the Fe$_\mathrm{B}$ sites and a smaller gap on the Fe$_\mathrm{A}$ sites. We can show this explicitly by calculating the density of states on the two sublattices as shown in Fig.~\ref{fig:dos}(b). We can clearly see that the gap size, as determined by the positions of the coherence peaks, is drastically changed between the sublattices. The parameters for this calculation were chosen so that the change in the SC gap between the sublattices is consistent with the largest gap modulation observed in Ref.~\cite{kongCooperpairDensityModulation2025}. In that experiment, the low-temperature measurement revealed a minimum coherence peak position of 1.06 meV and a maximum at 1.58 meV, which we can reproduce with the current model.

Our model, however, allows for more cases beyond the one discussed above, which lead to different behavior of the superconducting gap. For example, if both intra- and interband interactions are attractive, the signs of the order parameters will be the same for both bands, which means that the region with the largest gap magnitude will coincide with the same section of the Fermi surface for both bands. If the sublattice texture remains the same as previously discussed, this now means that the bands will exhibit a gap maximum at the opposite sublattice sites. If the experiment has a sufficient resolution, this will result in gap features that oscillate out of phase between the sublattices. Precise measurements of the behavior of the coherence peaks at the two sublattices can thus reveal information about the order parameter sign.

\begin{figure}
    \centering
    \includegraphics[width=0.99\columnwidth]{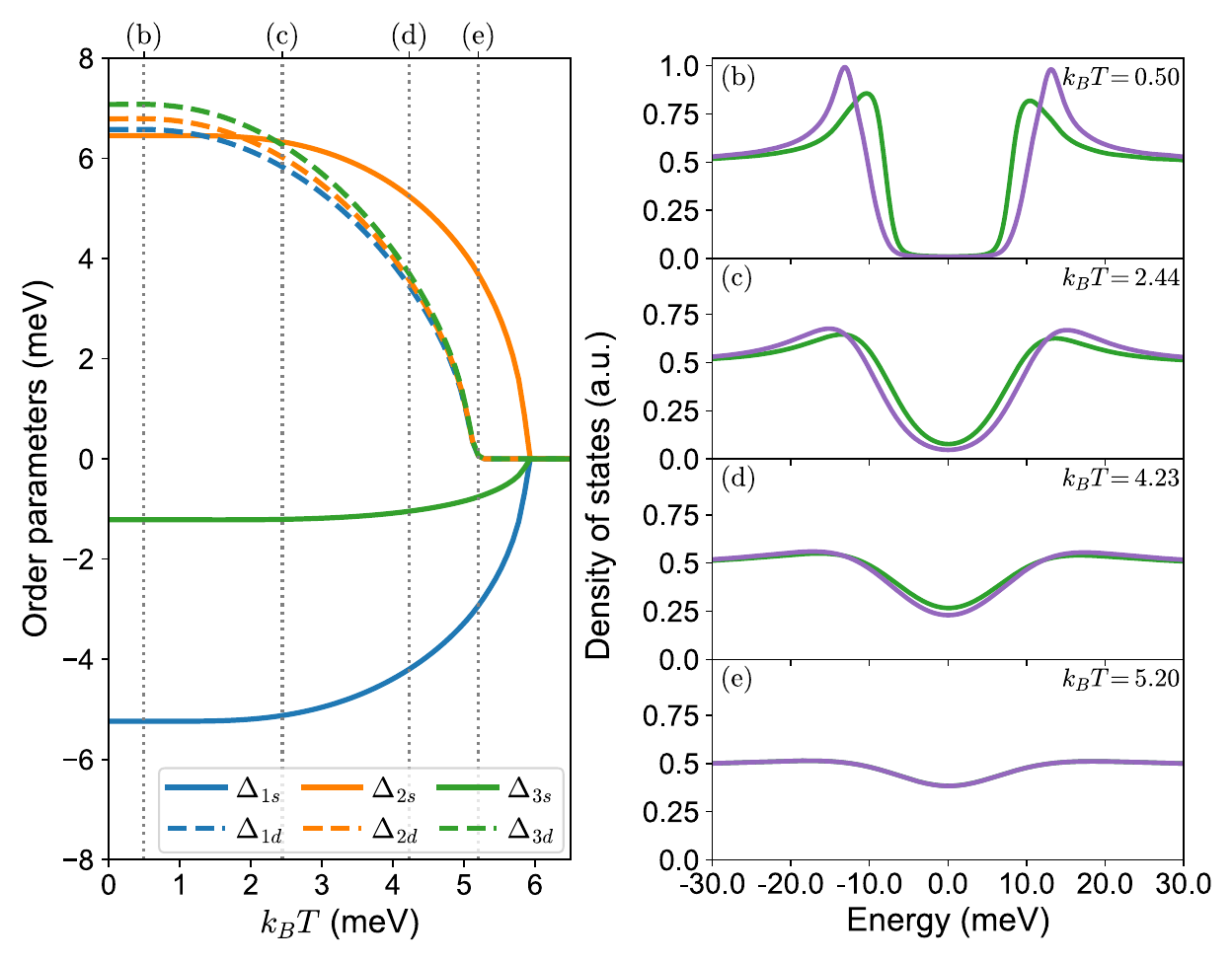}
    \caption{(a) Temperature dependence of all the order parameters for the three bands close to the Fermi energy. For this choice of interaction strengths ($V_{1s}=0.06, V_{1d}=0.4, V_{2s}=-0.03, V_{2d}=0.34$), the $d$ order parameter vanishes at a lower temperature, enabling existence of SC state with only $s$ pairing. The vertical lines indicate temperatures for which panels in right column are calculated. (b-e) Temperature evolution of the density of states at the two sublattices, which shows that the difference between Fe$_\mathrm{A}$ and Fe$_\mathrm{B}$ sites disappears when systems enters a pure $s_\pm$ phase.}
    \label{fig:temperature}
\end{figure}

\textit{Temperature dependence of PDM}.---
Another aspect of the pair density modulation is its dependence on temperature. In general, the two components of the mixed phase do not necessarily have to appear at the same temperature, leading to a situation where at the initial SC $T_c$ only a single component order parameter exists. We can thus have a purely $s_\pm$ or $d$ wave superconductor, and only after further cooling can the other component appear. This case is presented in Fig.~\ref{fig:temperature}(a), where initially only an $s_\pm$ phase appears. Due to the interband interactions, the $s_\pm$ order parameters for each of the bands arise at the same $T_c$. However, the $d$ order parameters emerge only at a lower temperature through a second-order phase transition. Both types of order parameters then increase towards their zero-temperature value. This is one type of behavior that we observed, but situations with non-monotonic behavior of the order parameters are also possible. In general, there is some competition between both order parameters, which can, for example, lead to an increase in the magnitude of the $d$ order parameter close to the transition temperature of the $s_\pm$ order parameter, compared to its value at $T=0$. In some circumstances, the mixed phase may exist only at intermediate temperatures, and not at $T=0$ at all.

Nevertheless, in the context of PDM, the most crucial aspect is that only the coexistence of the $s$ and $d$ order parameters can lead to a modulation in the gap size between the sublattices. Neither the pure $s_\pm$ nor the pure $d$ phase will result in a different gap at the Fe$_\mathrm{A}$ and Fe$_\mathrm{B}$ sites. This is because in the pure phases, there is no anisotropy of the order parameter magnitude that would line up with the sublattice texture. Therefore, when one of the order parameters disappears as the temperature increases, so will the gap modulation.

The consequences of a transition to a pure phase from a mixed phase are shown directly in the density of states at the two sublattices in a sequence of panels in Fig.~\ref{fig:temperature}(b-e). At the lowest temperature, the system is deep in the mixed phase, and so the density of states shows a clear difference in the superconducting gap at the two iron sites. However, as the temperature increases, the $d$ order parameter decreases more rapidly due to its lower transition temperature, and so the modulation between Fe$_\mathrm{A}$ and Fe$_\mathrm{B}$ becomes less apparent. Finally, above $d$ order $T_c$ the modulation completely disappears, yet the superconducting gap is still present due to the non-zero $s$ order parameter. Therefore, pair density modulation can allow for a direct visualization and study of the transition to a mixed two-component superconducting phase using local probe techniques.

\textit{Discussion}.---
In the preceding sections, we demonstrated that the combination of the sublattice texture that arises due to the glide-mirror symmetry breaking and the appearance of a mixed $s_\pm+d$ superconducting order due to interband interactions leads to the emergence of pair density modulation. In such a phase, the superconducting gap is spatially varying but still preserves the lattice translation symmetry, in contrast with other pair density wave states. Both ingredients are essential to this phenomenon: if there was no sublattice texture, the mixed superconducting phase would be spatially uniform, and in the pure $s$ or $d$ phases, there is no order parameter anisotropy that would align with the Fermi surface segments localized at either iron site. The same reasoning also applies if the orientation of the $d$ order parameter is not aligned with the sublattice texture: in the current case, if it was a $d_{xy}$ order instead of $d_{x^2-y^2}$, the modulation would not appear. This combination of factors may explain why pair density modulation was not observed earlier even though evidence for $s+d$ pairing was presented for some iron-based superconductors \cite{liNematicSuperconductingState2017, liuOrbitalOriginExtremely2018}. 
The key reason why PDM was seen only in exfoliated flakes may be that the exfoliation process strongly affects the layer spacing of the material and enhances the sublattice texture associated with breaking of glide-mirror symmetry. Moreover, the Fermi surface structure in exfoliated flakes is different than in the bulk samples (the M pockets are not present at the Fermi energy), which could also favor the appearance of an order parameter with $d_{x^2-y^2}$ symmetry.

Another important factor for why PDM was not detected in other systems is that the modulation here occurs on a very short length scale within the crystalline unit cell (much shorter than the superconducting coherence length commonly observed in iron-based superconductors), which requires extraordinary experimental effort to be measured. 
However, experiments can often be less focused on measurements of SC properties within the unit cell because conventional assumptions suggest that any changes in the superconducting order parameters should occur slowly on length scales comparable to the superconducting coherence length \cite{tinkhamIntroductionSuperconductivitySecond2015}. 
Our proposal sidesteps this issue as well, since the modulation here is unlike that of a pair density wave with its own separate finite-momentum order parameter. Here we have $s$ and $d$ order parameters, which on their own have longer coherence lengths, and only their combination together with the sublattice texture gives rise to short-period oscillation of the gap. This also means that our proposal is not susceptible to one possible issue with the microscopic mechanisms for pair density wave state, where the finite-momentum order parameter is strongly competing with zero-momentum pairing superconductivity, hindering their coexistence. However, in most of the observed PDW states, the gap modulation is a small contribution on top of a spatially uniform superconducting gap, and gapless excitations resulting from Fermi arcs are not observed in the tunneling spectra. In our theory for PDM, the exact behavior of the gap is determined by the relative magnitude of the $s$ and $d$ order parameters, but their existence requires only a much more common zero-momentum pairing. For relatively small $d$ order parameters, the nodal line does not cross the Fermi surface, and therefore the SC gap does not close, as presented in Fig.~\ref{fig:dos}(b). On the other hand, when the $d$ order parameter dominates, the coherence peak structure will still oscillate between sublattices, even though the SC gap will close at the nodal points.

The mechanism presented here also provides a possible resolution to the question of whether the pair density modulation state is a primary or a secondary order. As the density of states is spatially uniform above the superconducting transition temperature, there is no underlying charge density wave that induces the SC gap modulation - it arises purely from the nature of the superconducting state. Although here PDM is not described by a single order parameter, but as a combination of multiple factors, the nature of the superconducting pairing is still the central ingredient, independent of the precise properties of the normal state. The emergence of nematic $s+d$ superconductivity could also possibly serve as catalyst for the appearance of nematicity in the normal state dispersion, evidence of which was observed experimentally. Moreover, the separate transition temperature for PDM state is consistent with the temperature-dependent magnitude of the modulation Fourier component observed in the experiment \cite{kongCooperpairDensityModulation2025}. All of this supports pair density modulation as a primary order in the exfoliated flakes.

\textit{Summary and outlook}.---
In this work we presented a mechanism for a pair density modulation based on a combination of glide-mirror symmetry breaking and the appearance of a mixed $s+d$ superconducting phase. Future research can focus on elucidating what information about the superconducting pairing mechanism can be extracted from the gap structure oscillation within the crystalline unit cell, such as the connection with the sign of the order parameters and visualization of the phase transition between multi-component superconducting orders discussed in this work. However, the effects of sublattice texture should not be limited to superconducting properties. It may be worthwhile to explore other possible correlated ground states that arise in iron-based compounds, such as various magnetic phases, to see how they interact with the sublattice texture. The availability of additional information about the interacting phases through local probes may thus enable the investigation of the mechanisms behind their appearance. It also remains to be seen whether our proposed mechanism is relevant to other recently observed modulated superconducting states \cite{weiObservationSuperconductingPair2025, zhangVisualizingUniformLatticescale2024}. We also hope that our work will encourage more experimental groups to carefully investigate other materials, in the iron-based compound family and beyond, on the ultrashort length scales, revealing previously overlooked signals of pair density modulation and other interacting states.

\begin{acknowledgements}

\end{acknowledgements}

\bibliography{PDM_theory}

\newpage
\onecolumngrid
\setcounter{equation}{0}
\setcounter{figure}{0}
\setcounter{page}{1}
\renewcommand{\thefigure}{S\arabic{figure}}
\renewcommand{\theequation}{S\arabic{equation}}

\begin{center}
\textbf{\large Supplemental Material for ``Pair density modulation from glide symmetry breaking and nematic superconductivity"}

\,

{\normalsize Micha{\l} Papaj,$^1$ Lingyuan Kong,$^{2,3}$ Stevan Nadj-Perge,$^{2,3}$ Patrick A. Lee$^4$}

\,\\

\textit{\small $^1$Department of Physics and Texas Center for Superconductivity\\ at the University of Houston (TcSUH), Houston, TX 77204, USA \\
$^{2}$T. J. Watson Laboratory of Applied Physics, California Institute of Technology,\\ 1200 East California Boulevard, Pasadena, California 91125, USA \\
$^{3}$Institute for Quantum Information and Matter,\\ California Institute of Technology, Pasadena, California 91125, USA\\
$^4$Department of Physics, Massachusetts Institute of Technology, Cambridge, MA 02139, USA
}

\author{Lingyuan Kong}
\affiliation{T. J. Watson Laboratory of Applied Physics, California Institute of Technology, 1200 East California Boulevard, Pasadena, California 91125, USA}
\affiliation{Institute for Quantum Information and Matter, California Institute of Technology, Pasadena, California 91125, USA}

\author{Stevan Nadj-Perge}
\affiliation{T. J. Watson Laboratory of Applied Physics, California Institute of Technology, 1200 East California Boulevard, Pasadena, California 91125, USA}
\affiliation{Institute for Quantum Information and Matter, California Institute of Technology, Pasadena, California 91125, USA}

\author{Patrick A. Lee}
\affiliation{Department of Physics, Massachusetts Institute of Technology, Cambridge, MA 02139, USA}
\end{center}

\section{Tight-binding model with glide mirror symmetry broken}

We derive the tight-binding model for the two-iron unit cell of Fe(Se, Te) based on the Koster-Slater (KS) approach using $d_{xz}$, $d_{yz}$, $d_{xy}$ orbitals on Fe atoms and $p_x$, $p_y$, $p_z$ orbitals on Te/Se atoms. The atomic positions are as follows:
\begin{equation}
    \text{Fe}_\text{A}: (0, 0, 0), \quad \text{Fe}_\text{B}: \left(\frac{a}{2}, \frac{a}{2}, 0\right), \quad \text{Se}_\text{A}: \left(\frac{a}{2}, 0, c_1\right), \quad \text{Se}_\text{B}:  \left(0, \frac{a}{2}, -c_2\right).
\end{equation}
The lattice vectors are ($a$, 0, 0) and (0, $a$, 0), where $a$ is the distance between the iron atoms that belong to the same sublattice (i.e. the next-nearest neighbors). To break the glide mirror symmetry we allow $c_1 \neq c_2$ and so all the KS integrals (which depend on the distance between atoms) between Fe atoms and Se/Te atoms are different for Se/Te atoms in the upper and lower plane. In this intermediate step, we obtain eight independent parameters that describe nearest-neighbor Fe-Se hoppings between the $d$ and $p$ orbitals (this includes hoppings that differ due to $c_1 \neq c_2$) and five parameters that describe hoppings between $d$ orbitals for nearest and next-nearest-neighbor iron atoms. We can therefore obtain a lattice Hamiltonian $H_\text{lat} = H_\text{diag} + H'$, where $H_\text{diag}$ are the onsite energies of different orbitals ($\epsilon_p$ for $p_x$ and $p_y$, $\epsilon_{p_z}$ for $p_z$, $\epsilon_d$ for $d_{xz}$ and $d_{yz}$, and $\epsilon_{d_{xy}}$ for $d_{xy}$) and $H'$ are the hoppings from the KS procedure.

Since we are interested in the low energy properties of Fe electrons, we treat the Se/Te orbitals in the second order of perturbation theory and include their impact on Fe states through the Schrieffer-Wolff transformation:
\begin{gather}
    H^\text{SW}_{mm'} = H^{(0)}_{mm'} + H^{(1)}_{mm'} + H^{(2)}_{mm'}, \\
    H^{(0)}_{mm'} = H_{\text{diag},mm'}, \quad H^{(1)}_{mm'} = H'_{mm'}, \quad H^{(2)}_{mm'} = \frac{1}{2} \sum_l H'_{ml}H'_{lm'} \left[ \frac{1}{E_m - E_l} + \frac{1}{E_{m'} - E_l} \right].
\end{gather}
In the equations above, $m$ indexes the $d$ orbital states on the iron atoms and $l$ indexes the $p$ orbital states on the Se/Te atoms. The next step is to perform Fourier transform on the lattice Hamiltonian $H^\text{SW}$ to finally obtain the tight-binding Hamiltonian in momentum space:
\begin{equation}
    H_\text{TB}(\mathbf{k}) = 
    \begin{pmatrix}
    \epsilon_{11AA} & 0 & \epsilon_{13AA} & \epsilon_{11AB} & \epsilon_{12AB} & \epsilon_{13AB} \\
    0 & \epsilon_{22AA} & \epsilon_{23AA} & \epsilon_{21AB} & \epsilon_{22AB} & \epsilon_{23AB} \\
    \epsilon_{13AA}^* & \epsilon_{23AA}^* & \epsilon_{33AA} & \epsilon_{31AB} & \epsilon_{32AB} & \epsilon_{33AB} \\
    \epsilon_{11AB} & \epsilon_{12AB} & \epsilon_{31AB}^* & \epsilon_{11BB} & 0 & \epsilon_{13BB} \\
    \epsilon_{12AB} & \epsilon_{22AB} & \epsilon_{32AB}^* & 0 & \epsilon_{22BB} & \epsilon_{23BB} \\
    \epsilon_{13AB}^* & \epsilon_{23AB}^* & \epsilon_{33AB} & \epsilon_{13BB}^* & \epsilon_{23BB}^* & \epsilon_{33BB} \\
    \end{pmatrix}
\end{equation}
with the matrix elements defined as follows:

\begin{align}
    \epsilon_{11AA} = \epsilon_{d1} + 2(t_{5a} \cos k_x a + t_{6b} \cos k_y a), \quad \epsilon_{13AA} = 2i t_{7b} \sin k_y a, \quad \epsilon_{11AB} = \epsilon_{22AB} = 4 t_1 \cos\frac{k_x a}{2}\cos\frac{k_y a}{2}, \\ \epsilon_{12AB} = 4 t_{2a} \sin\frac{k_x a}{2}\sin\frac{k_y a}{2}, \quad \epsilon_{13AB} = -4i t_{4a} \cos\frac{k_x a}{2}\sin\frac{k_y a}{2}, \quad \epsilon_{22AA} = \epsilon_{d2} + 2(t_{6a} \cos k_x a + t_{5b} \cos k_y a),  \\ \epsilon_{23AA} = -2i t_{7a} \sin k_x a, \quad \epsilon_{21AB} =  4 t_{2b} \sin\frac{k_x a}{2}\sin\frac{k_y a}{2}, \quad \epsilon_{23AB} = 4i t_{4b} \sin\frac{k_x a}{2}\cos\frac{k_y a}{2}, \\
    \epsilon_{33AA} = \epsilon_{d3} + 2 (t_{8a} \cos k_x a + t_{8b} \cos k_y a), \quad \epsilon_{31AB} = -4i t_{4b} \cos\frac{k_x a}{2}\sin\frac{k_y a}{2}, \quad \epsilon_{32AB} = 4i t_{4a} \sin\frac{k_x a}{2}\cos\frac{k_y a}{2}
\end{align}
\begin{align}
    \epsilon_{33AB} = 4 t_{3} \cos\frac{k_x a}{2}\cos\frac{k_y a}{2}, \quad \epsilon_{11BB} = \epsilon_{d2} + 2(t_{5b} \cos k_x a + t_{6a} \cos k_y a), \quad \epsilon_{13BB} = -2i t_{7a} \sin k_y a, \\ \epsilon_{22BB} = \epsilon_{d1} + 2(t_{6b} \cos k_x a + t_{5a} \cos k_y a), \quad \epsilon_{23BB} = 2i t_{7b} \sin k_x a, \quad
    \epsilon_{33BB} = \epsilon_{d3} + 2 (t_{8b} \cos k_x a + t_{8a} \cos k_y a)  
\end{align}
where $t_{ia/b}$ are the hopping parameters, with $a/b$ distinguishing parameters that would be equal in the absence of glide mirror symmetry breaking, and $\epsilon_{di}$ are the onsite energies of different orbitals modified by the perturbations obtained through the Schrieffer-Wolff transformation. While this Hamiltonian includes three $d$ orbitals of iron atoms, since we are focusing on the $\Gamma$ point pockets, to simplify the model, we drop the $d_{xy}$ components. After doing so, we perform a unitary transformation:
\begin{equation}
    \tilde{H}_\text{TB} = U H_\text{TB} U^\dagger, \quad U = \frac{1}{\sqrt{2}} \tau_0 \otimes \begin{pmatrix} 1 & 1 \\ -1 & 1 \end{pmatrix}
\end{equation}
where $\tau_0$ is the Pauli matrix in the sublattice subspace. After this transformation, we perform the Taylor expansion of the Hamiltonian close to $k_x = k_y = 0$ point. When we express momenta in units of $1/a$, we obtain:
\begin{equation}
\label{eq:SM_ham_cont}
    H_0(\mathbf{k}) = 
    \begin{pmatrix}
        \epsilon_{\text{A}} & \epsilon_{\text{A},12} & \epsilon_{\text{AB},1} & \epsilon_{\text{AB},12} \\
        \epsilon_{\text{A},12} & \epsilon_{\text{A}} & -\epsilon_{\text{AB},12} & \epsilon_{\text{AB},2} \\
        \epsilon_{\text{AB},1} & -\epsilon_{\text{AB},12} & \epsilon_{\text{B}} & \epsilon_{\text{B},12} \\
        \epsilon_{\text{AB},12} & \epsilon_{\text{AB},2} & \epsilon_{\text{B},12} & \epsilon_{\text{B}}
    \end{pmatrix}
\end{equation}
with $\epsilon_{\text{A/B}} = -\alpha k^2 \mp \beta (k_x^2-k_y^2) - \mu$, $\epsilon_{\text{AB},1/2} = \xi (k^2 - 8) \pm \rho k_x k_y$, $\epsilon_{\text{A/B},12}= \pm \delta + \gamma (k_x^2-k_y^2) \pm \kappa (k^2 - 4)$, $\epsilon_{\text{AB},12}=\chi k_x k_y$, which is the Hamiltonian presented in the main text. We also include a parabolic band $\epsilon_M = \epsilon_0 + \nu k^2 - \mu$ that aims to reproduce the electron pocket visible in the quasiparticle interference experiments at energies above the Fermi level. We assume that this band is located at the M point, consistent with the spectrum of other iron-based superconductors.

In the calculations, we use the following values of the parameters $\alpha = 0.3$, $\beta = 0.0$, $\xi = 0.05$, $\rho = 0.05$, $\delta = 0.1$, $\gamma = -0.075$, $\kappa = -0.1$, $\chi=0.0$, $\nu = 0.2$, with the energies expressed in eV. Although this is one possible set of parameters, the character of the superconducting order parameters obtained through self-consistent calculations described in the next section is not strongly dependent on the exact values. However, the sublattice texture is strongly affected by the choice of parameters. For example, if all the parameters $\beta$, $\delta$, $\kappa$, and $\chi$ are zero, the glide mirror symmetry is preserved and the sublattice texture is zero for both bands of interest.

\begin{figure}[b]
    \centering
    \includegraphics[width=0.336\linewidth]{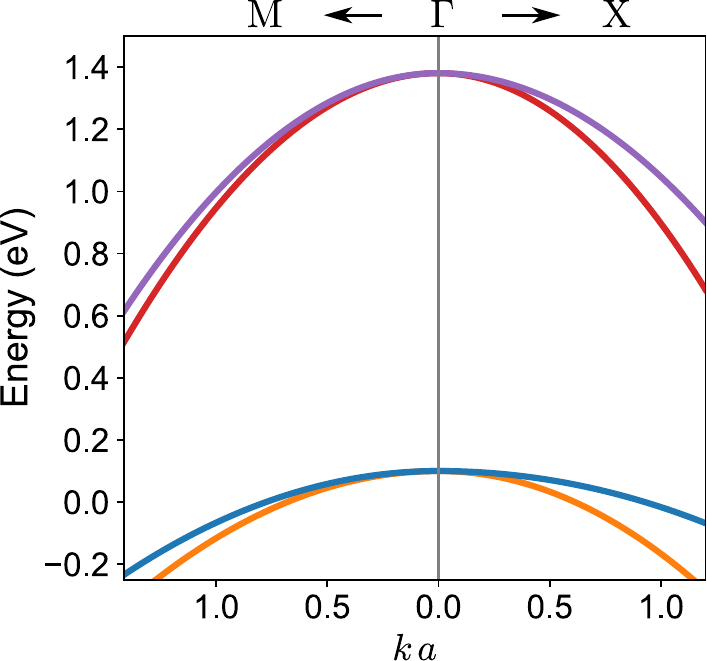}
    \caption{The full spectrum of Hamiltonian \eqref{eq:SM_ham_cont}, which includes four bands that form two pairs degenerate at $\Gamma$ point. We disregard the two higher energy bands when considering superconducting pairing as they are not present at the Fermi energy.}
    \label{fig:SM_full_spectrum}
\end{figure}

The full spectrum of this model is presented in Fig.~\ref{fig:SM_full_spectrum}, which displays all four bands, split into a higher and a lower energy doublets. Since we are only interested in the bands that form the Fermi surface or are very close to the Fermi energy, we disregard the two higher energy bands, focusing on the two lower ones as discussed in the main text.

The top of the two bottom bands is located at $-\sqrt{64 \xi^2 + (\delta - 4\kappa)}-\mu$. We choose $\mu$ so that the Fermi level is placed at 0.1 eV below the top of these bands for visualization of the band dispersion, the sublattice texture, the phase diagram calculations, and temperature-dependent calculations. For the example shown in Fig. 3 of the main text that aims to reproduce the experimental results, we set the Fermi level 0.01 eV below the top of the hole bands, in accordance with estimation from quasiparticle interference.

\section{Self-consistent solutions of the gap equations}
To determine the phase diagram of the superconducting states in the studied model, we first diagonalize the Hamiltonian of Eq.~\eqref{eq:SM_ham_cont}, obtaining four bands split into two doublets, one of which lies close to the Fermi energy. We take these two lowest-energy bands at the $\Gamma$ point and also include the parabolic band $\epsilon_M$. We label the dispersions of these three bands $\epsilon_i(\mathbf{k})$. Within this subspace, we consider interactions as discussed in the main text:
\begin{equation}
    H_\text{int} = -\frac{1}{N} \sum_{\mathbf{k}, \mathbf{k'}} \left[  V_1(\mathbf{k}, \mathbf{k'}) \sum_i c^\dagger_{i\mathbf{k}\uparrow} c^\dagger_{i-\mathbf{k}\downarrow} c_{i-\mathbf{k'}\downarrow} c_{i\mathbf{k'}\uparrow} +   V_2(\mathbf{k}, \mathbf{k'}) \sum_{i\neq j} c^\dagger_{i\mathbf{k}\uparrow} c^\dagger_{i-\mathbf{k}\downarrow} c_{j-\mathbf{k'}\downarrow} c_{j\mathbf{k'}\uparrow} \right]
\end{equation}
where $V_n(\mathbf{k}, \mathbf{k'}) = V_{ns}f_s(\mathbf{k})f_s(\mathbf{k'}) + V_{nd}f_d(\mathbf{k})f_d(\mathbf{k'})$ and $f_s(\mathbf{k}) = \cos(k_x) + \cos(k_y)$, $f_d(\mathbf{k}) = \cos(k_x) - \cos(k_y)$ are the $s_\pm$ and $d$ channel interaction form factors, respectively. We also keep in mind that the parabolic band $\epsilon_M$ (now labeled $\epsilon_3(\mathbf{k})$) is located at the M point, so in that case $\mathbf{k}$ in the interaction form factors is understood to mean $\mathbf{k}+\mathbf{Q}$ with $\mathbf{Q}=(\pi, \pi)$. We can now introduce the $s_\pm$ and $d$ order parameters for each of the bands considered in the model:
\begin{equation}
    \Delta^{(i)}(\mathbf{k}) = \Delta_{s}^{(i)} f_s(\mathbf{k}) + \Delta_{d}^{(i)} f_d(\mathbf{k})
\end{equation}
\begin{equation}
    \Delta^{(i)}_s = -\frac{1}{N} \sum_{\mathbf{k'}} \left[ V_{1s} f_s(\mathbf{k'}) \langle c_{i-\mathbf{k'}\downarrow} c_{i\mathbf{k'}\uparrow} \rangle + V_{2s}f_s(\mathbf{k'}) \sum_{j\neq i} \langle c_{j-\mathbf{k'}\downarrow} c_{j\mathbf{k'}\uparrow} \rangle \right]
\end{equation}
\begin{equation}
    \Delta^{(i)}_d = -\frac{1}{N} \sum_{\mathbf{k'}} \left[ V_{1d} f_d(\mathbf{k'}) \langle c_{i-\mathbf{k'}\downarrow} c_{i\mathbf{k'}\uparrow} \rangle + V_{2d}f_d(\mathbf{k'}) \sum_{j\neq i} \langle c_{j-\mathbf{k'}\downarrow} c_{j\mathbf{k'}\uparrow} \rangle \right]
\end{equation}
The expectation values are calculated using standard Matsubara Green function approach in Nambu space. We first define a Bogoliubov-de Gennes Hamiltonian as follows:
\begin{equation}
    H^\text{(b)}_\text{BdG}(\mathbf{k}) = \begin{pmatrix}
        H_\text{b}(\mathbf{k}) & \Delta(\mathbf{k}) \\
        \Delta(\mathbf{k})^\dagger & -H_\text{b}(-\mathbf{k})
    \end{pmatrix},
    \, \, \, \,
    H_\text{b}(\mathbf{k}) = \begin{pmatrix}
        \epsilon_{1}(\mathbf{k}) & 0 & 0 \\
        0 & \epsilon_{2}(\mathbf{k}) & 0 \\
        0 & 0 & \epsilon_{3}(\mathbf{k})
    \end{pmatrix},
    \, \, \, \,
    \Delta(\mathbf{k}) = \begin{pmatrix}
        \Delta^{(1)}(\mathbf{k}) & 0 & 0 \\
        0 & \Delta^{(2)}(\mathbf{k}) & 0 \\
        0 & 0 & \Delta^{(3)}(\mathbf{k})
    \end{pmatrix},
\end{equation}
and then define the Green function as $G^\text{(b)}(i\omega_n, \mathbf{k}) = (i\omega_n - H^\text{(b)}_\text{BdG}(\mathbf{k}))^{-1}$. The expectation values are then Matsubara frequency sums of appropriate matrix elements of $G^\text{(b)}(i\omega_n, \mathbf{k})$.

The gap equations are solved self-consistently starting from a random initial condition. The phase diagram is established by choosing a cutoff point of $\Delta_c = 10^{-4}$ eV: if both $s_\pm$ and $d$ order parameters exceed $\Delta_c$, the phase is labeled mixed $s_\pm+d$, and when only one of the $s_\pm$ or $d$ order parameters exceeds $\Delta_c$, the phase is labeled pure $s_\pm$ or pure $d$.

To calculate the density of states at the two iron sublattices, we transform the Hamiltonian $H_\text{BdG}^\text{(b)}(\mathbf{k})$ with the superconducting order parameters obtained through the self-consistent procedure from the band basis back to the sublattice/orbital basis using the unitary transformation determined during the original diagonalization of $H_0(\mathbf{k})$ of Eq.~\eqref{eq:SM_ham_cont}. We find the retarded Green function for the Bogoliubov-de Gennes Hamiltonian in the sublattice/orbital space $H^\text{(l)}_\text{BdG}(\mathbf{k})$ as $G^{\text{(l)}}(\omega + i\eta, \mathbf{k}) = (\omega + i\eta - H^\text{(l)}_\text{BdG}(\mathbf{k}))^{-1}$. With that, we get the A/B sublattice-resolved density of states $n_A(\omega), n_B(\omega)$ as:
\begin{equation}
    n_A(\omega) = - \frac{1}{\pi} \text{Im} \sum_{\mathbf{k}} \left( G^{\text{(l)}}_{11}(\omega+i\eta, \mathbf{k}) + G^{\text{(l)}}_{22}(\omega+i\eta, \mathbf{k}) \right)
\end{equation}
\begin{equation}
    n_B(\omega) = - \frac{1}{\pi} \text{Im} \sum_{\mathbf{k}} \left( G^{\text{(l)}}_{33}(\omega+i\eta, \mathbf{k}) + G^{\text{(l)}}_{44}(\omega+i\eta, \mathbf{k}) \right)
\end{equation}

\begin{figure}
    \centering
    \includegraphics[width=0.999\linewidth]{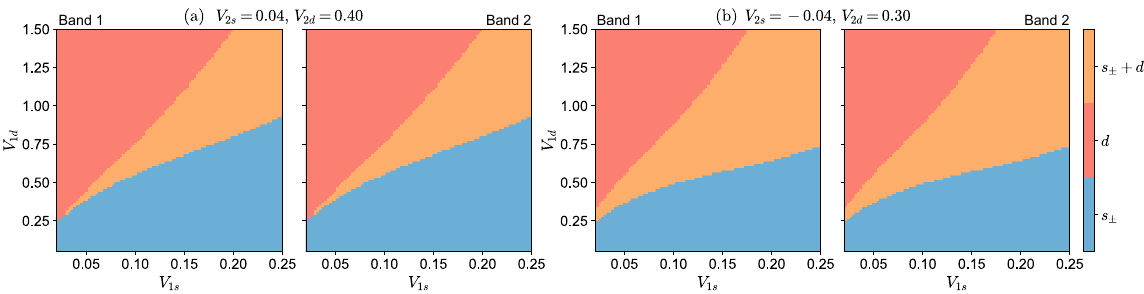}
    \caption{The zero temperature phase diagrams for two sets of interband interaction strengths: (a) attractive in $s_\pm$ channel and (b) repulsive in $s_\pm$ channel. In both cases, both bands exhibit a mixed $s+d$ phase in a wide range of parameters.}
    \label{fig:SM_both_bands}
\end{figure}

\section{Additional zero temperature phase diagrams}
In Fig.~\ref{fig:SM_both_bands} we present the zero-temperature phase diagrams for both bands for attractive and repulsive interband interactions, expanding the results shown in the main text. This shows that both bands are mostly exhibiting the same phase as a result of the presence of interband interactions, even though the exact values of the order parameters are different for each of them. The relative sign between the order parameters can also change, as shown in the main text. Because we allow for different order parameters in each of the bands, this in general can lead to a complex set of gap features, whose evolution within the unit cell will be governed by the shape of the Fermi surfaces, order parameter anisotropy, and the sublattice texture. Further exploration of this behavior can reveal more information about the pairing mechanisms and the nature of the superconducting state.

\begin{figure}
    \centering
    \includegraphics[width=0.53\linewidth]{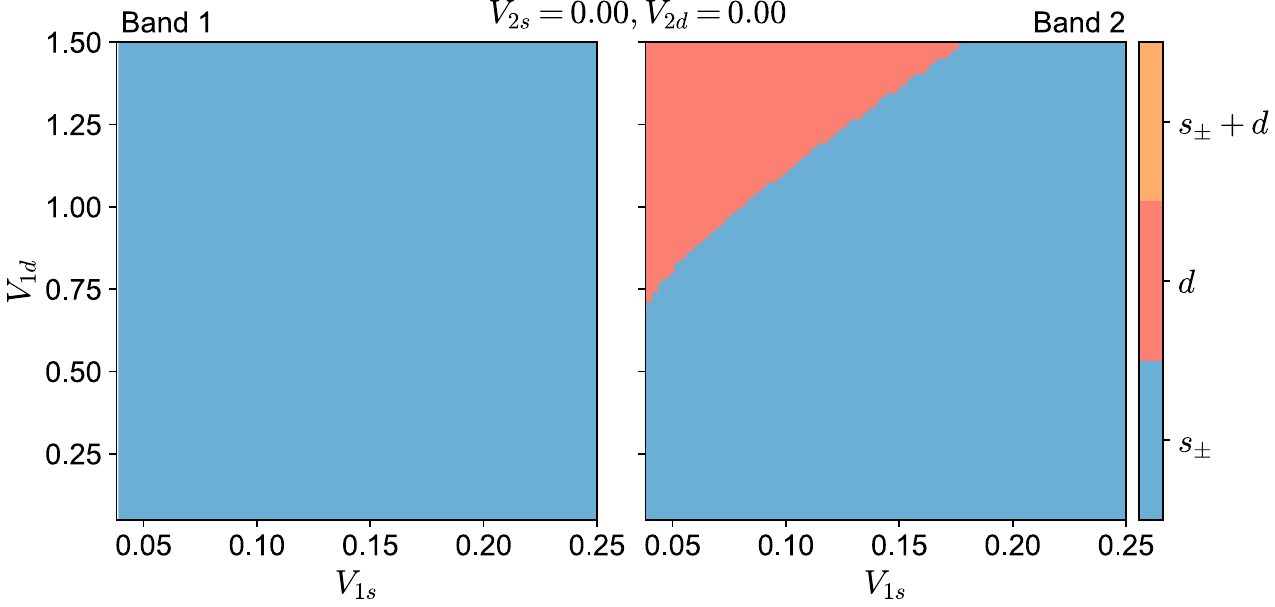}
    \caption{Zero temperature phase diagrams for both bands without any interband interaction. In the scanned parameter range, band 1 exhibits only an $s_\pm$ phase, while band 2 is split between pure $s_\pm$ and pure $d$ phase, without any mixed $s+d$ order.}
    \label{fig:SM_no_interband}
\end{figure}

In Fig.~\ref{fig:SM_no_interband} we show the zero-temperature phase diagrams for both bands in the case where there are no interband interactions. This demonstrates that in the absence of interband interactions, the mixed $s+d$ phase does not appear, and instead we have a first-order phase transition between the pure $s$ and pure $d$ orders. Also, without the interband interactions, in general the order parameters in two bands will exhibit a different behavior, since in the example shown, band 1 remains only in the $s$ phase in the studied interaction strength range, while band 2 can also enter the $d$ phase.

\end{document}